# Shape resonances in multi-condensate granular superconductors formed by networks of nanoscale-striped puddles

**Antonio Bianconi**

*RICMASS Rome International Center for Materials Science Superstripes, Via dei Sabelli 119A, 00185 Roma, Italy*

E-mail: antonio.bianconi@ricmass.eu

**Abstract.** A characteristic feature of a superconductor made of multiple condensates is the possibility of the shape resonances in superconducting gaps. Shape resonances belong to class of Fano resonances in configuration interaction between open and closed scattering channels. The Shape resonances arise because of the exchange interaction, a Josephson-like term, for transfer of pairs between different condensates in different Fermi surface spots in the special cases where at least one Fermi surface is near a 2.5 Lifshitz topological transition. We show that tuning the shape resonances show first, the gap suppression (like a Fano anti-resonance) driven by configuration interaction between a BCS condensate and a BEC-like condensate, and second, the gap amplification (like a Fano resonance) driven by configuration interaction between BCS condensates in large and small Fermi surfaces. Shape resonances usually occur in granular nanoscale complex matter (called superstripes) because of the lattice instability near a 2.5 Lifshitz transition in presence of interactions. Using a new imaging method, scanning nano-X-ray diffraction, we have shown the generic formation in high temperature superconductors of a granular superconducting networks made of striped puddles formed by ordered oxygen interstitials or ordered local lattice distortions (like static short range charge density waves). In the superconducting puddles the chemical potential is tuned to a shape resonance in superconducting gaps and the maximum $T_c$ occurs where the puddles form scale free superconducting networks.

## 1. Introduction

Understanding the mechanism that allows a quantum condensate to resist decoherence attacks of temperature is a major fundamental problem of condensed matter. The Bardeen Cooper Schrieffer wave-function [1-5] of the superconducting ground state has been constructed based on the theory of configuration interaction of all possible electron pairs (+k with spin up, and -k with spin down) on the Fermi surface in an energy window called the energy cut off of the interaction,

$$|\Psi_{BCS}\rangle = \prod_k (u_k + v_k c^+_{k\uparrow} c^+_{-k\downarrow})|0\rangle \qquad (1)$$

where $|0\rangle$ is the vacuum state, and $c_{k\uparrow}^+$ is the creation operator for an electron with momentum k and spin up. The many body BCS condensate with off-diagonal long range appears at the gap energy $\Delta(\kappa)$ below the Fermi level. The Schrieffer idea [5] came from the configuration interaction theory by Tomonaga involving a pion condensate around the nucleus [6]. The k-dependent structure of the interaction gives different values for the gap $\Delta(k)$ in different segments of the Fermi surface. The superfluid order parameter i.e., the superconducting gap is anisotropic in the k-space i.e., different in different locations of the k-space due different pairing strength. The k-dependent gap equation is given by

$$\Delta(\mu,k) = -\frac{1}{2N}\sum_{k'} \frac{V(k,k')\Delta(k_y')}{\sqrt{(E(k')+\varepsilon_{k'}-\mu)^2 + \Delta^2(k_y')}} \quad (2)$$

the critical temperature is given by

$$\Delta(k) = -\frac{1}{N}\sum_{k'} V(k,k') \frac{tgh(\frac{\xi(k')}{2T_c})}{2\xi(k')}\Delta(k') \quad (3)$$

where the $\xi_n(k) = \varepsilon_n(k) - \mu$

These original BCS formulas describe the anisotropic superconductivity in the "clean limit", where the single electron mean free path is larger than the superconducting coherence length. Real superconducting materials [7,8] have impurities and lattice disorder therefore the condition for the mean-free path $l > hv_F/\Delta_{av}$ where $v_F$ is the Fermi velocity and $\Delta_{av}$ is the average superconducting gap was considered to be impossible to be satisfied [9,10]. In fact it is a very strict condition that implies that the impurity scattering rate $\gamma_{ab} << (1/2)(K_B/\hbar)T_c$ i.e., it should be smaller than few meV. The 40 years period from 1960 to 2000 was dominated by the *"dirty limit" dogma* [9,10] assuming an effective single Fermi surface for each superconductor. This dogma stated that all metals are in the *"dirty limit"* since impurity scattering and hybridization always mixes the wave functions of electrons on different spots in the same Fermi surface or in different Fermi surfaces where there are multiple bands crossing the Fermi level. The *"dirty limit"* dogma justifies the approximation in the theory of a k-independent V(k,k') interaction i.e., an isotropic pairing interaction that is assumed to be a constant $V_0$ [9,10]. This approximation allows to derive the approximated simple universal BCS formula for the critical temperature $T_C$ related to a single superconducting energy gap $\Delta_0$ the energy needed to break the cooper pairs, averaged over many k-points in different bands:

$$\frac{2\Delta_0}{K_B T_C} = 3.52 \quad (4)$$

$$\frac{T_c}{T_F} = \frac{0.36}{k_F \xi_0} \quad (5)$$

$$K_B T_c \propto \hbar\omega_0 e^{-1/\lambda_{eff}} \quad (6)$$

where $T_F$ is Fermi temperature, $k_F=2\pi/\lambda_F$ is the wave-vector of electrons at the Fermi level, $\xi_0$ is the coherence length of the condensate, related with the size of the pair, $\hbar\omega_0$ is the energy if the boson meditating the pairing interaction, the effective coupling term $\lambda_{eff} = V_o N_{tot}$ is the product of the pairing strength $V_0$ and $N_{tot}$ is the total density of states (DOS) at the Fermi level. Moreover using this approximation the BCS theory predicts that the critical temperature depends on isotope substitution with power law dependence $T_c \propto M^{-\alpha}$ where M is the atomic mass and α=0.5 is the isotope coefficient [10].

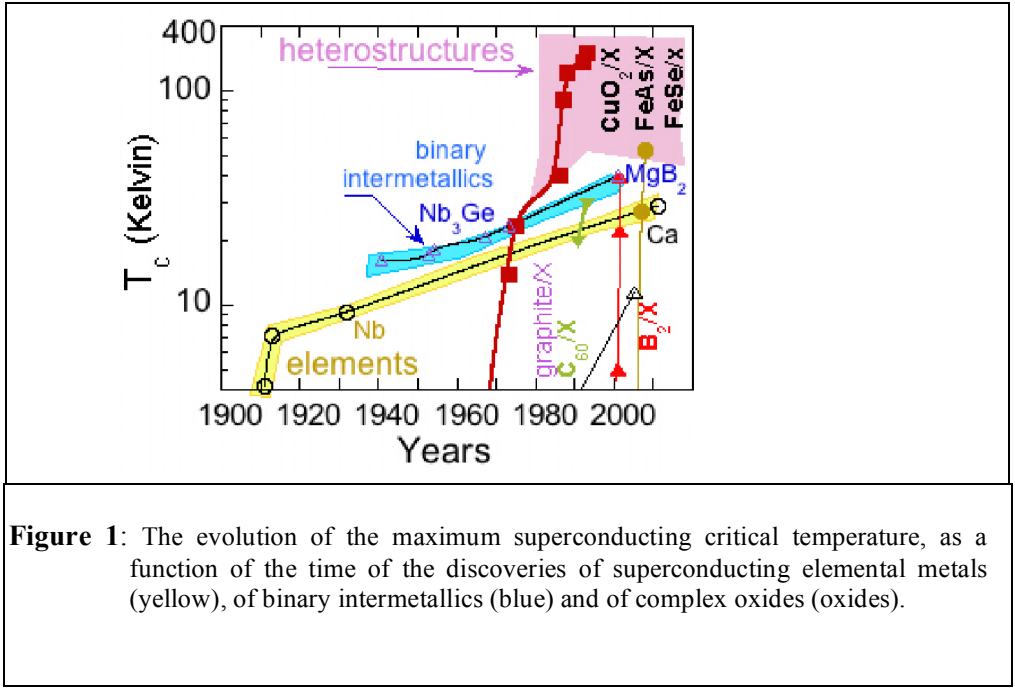

**Figure 1**: The evolution of the maximum superconducting critical temperature, as a function of the time of the discoveries of superconducting elemental metals (yellow), of binary intermetallics (blue) and of complex oxides (oxides).

These popular formulas are simple but they have completely washed out all interesting possible effects on the macroscopic superconducting parameters due to quantum interference effects in the configuration interaction between pairing channels in different points of the k-space that could appear in the *"clean limit"*. According with De Gennes this is not a problem since no spectacular effects have to be expected from the k-dependent V(k,k') interaction [10]. The last 40 years of the XX century can be called the period of the *"dirty limit"* dogma. The road map for the research of new high $T_c$ materials based on the BCS approximated formulas was focused on metals showing peaks of total DOS, strong electron phonon coupling, and high energy phonon modes. The details of superconductivity in different materials have been described by the Migdal-Eliashberg approximation including the details if the attractive electron-phonon interaction and the Coulomb repulsion, keeping the assumption that the energy scale of the pairing interaction is much smaller than the Fermi energy $\hbar\omega_0 << E_F$ called the so called *"adiabatic limit"* i.e. the chemical potential in the metallic material is assumed to be far away from band-edge. The materials enter in the anti-adiabatic regime $E_F \approx \hbar\omega_0$ and the charge carriers become large or small polarons. Moving the chemical potential toward a band edge where the system approaches a metal-to-insulator transition. In the extreme antiadiabatic regime $E_F << \hbar\omega_0$ depending on the coupling one can have all possibilities from weak coupling BCS to BEC (Bose Einstein Condensate) that could be also below the band edge. Therefore near a band edge the system could be in the so called BEC-BCS crossover [11].

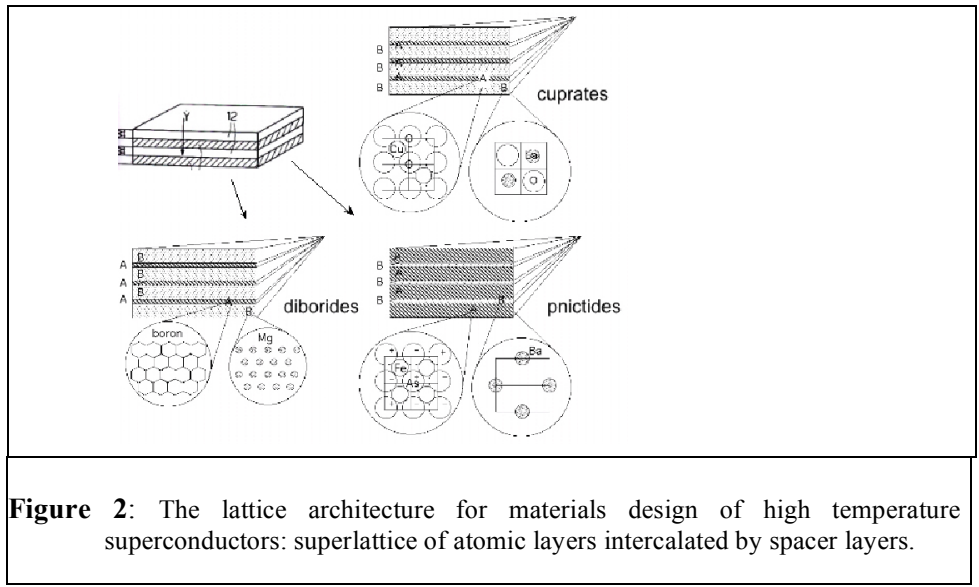

**Figure 2**: The lattice architecture for materials design of high temperature superconductors: superlattice of atomic layers intercalated by spacer layers.

The slow progress of materials science in the search for new high temperature superconductors in these last 100 years is shown in Fig. 1. The first discontinuity occurred by shifting from simple elemental metals to binary intermetallics in the 30's and the record was reached in metastable phases near a lattice instability. The second discontinuity in 1986 was the research shift from intermetallics to ceramics. In 1993 it was recognized that the cuprates are heterostructures at atomic limit therefore new high temperature superconductors should be made of superlattices of atomic superconducting units intercalated by spacers as the superlattice of superconducting atomic layers shown in Fig. 2 [12-15]. This theory has been confirmed in these last 20 years since all high temperature superconductors discovered so far are all made of similar heterostructures at atomic limit: diboride atomic layers [16-19], superlattices of iron based pnictides and chalcogenides layers [20-22] graphene, nanotubes [23] organic units intercalated by spacers. In this scenario the high $T_c$ is controlled by the shape resonance [12-18] in superconducting gaps where one band is near an electronic topological transitions, that is a 2 1/2-order transition, called 2.5 Lifshitz transition [24-25].

The 2.5 Lifshitz transition was first introduced by Lifshitz in 1961 [24] and has been widely studied in solid state physics. The kinetic properties of metals at a 2.5 Lifshitz transition are well known [25-26]. The generic feature of the Fermi surface topology near a band edge in a multiband metal made of a superlattice of atomic layers is shown in Fig. 3. Two types of topological 2.5 Lifshitz transitions [18] appears by tuning the chemical potential near a band edge.

A clear practical realization of this scenario is $MgB_2$, a material known since 1953, where superconductivity was never measured for 46 years since no superconductivity was predicted by standard theories based on the effective single band "dirty limit" dogma but on the contrary high temperature superconductivity was predicted by the multiband theory if the chemical potential is driven near a band edge i.e., below the top of the boron $2p_{x,y}$ sigma band where it should form a multi-gap superconductor in the "clean limit" because of single electron hopping between π and σ bands are forbidden by symmetry [16-19]. The discovery of high $T_c$ superconductivity in this material has shown that high temperature superconductivity appears where both the *"dirty limit dogma"* and the *adiabatic approximation* in BCS theory breakdown as first recognized in 2001 [16]. In fact the adiabatic condition breaks down near a 2.5 Lifshitz transition. However the anisotropic BCS theory [11] may works in the anti-adiabatic limit in multiband superconductors. This is shown by the fact that BCS seems to work pretty quantitatively for the Feshbach resonance in ultracold gases and for fullerenes despite the fact that the adiabatic condition is not at all well fulfilled.

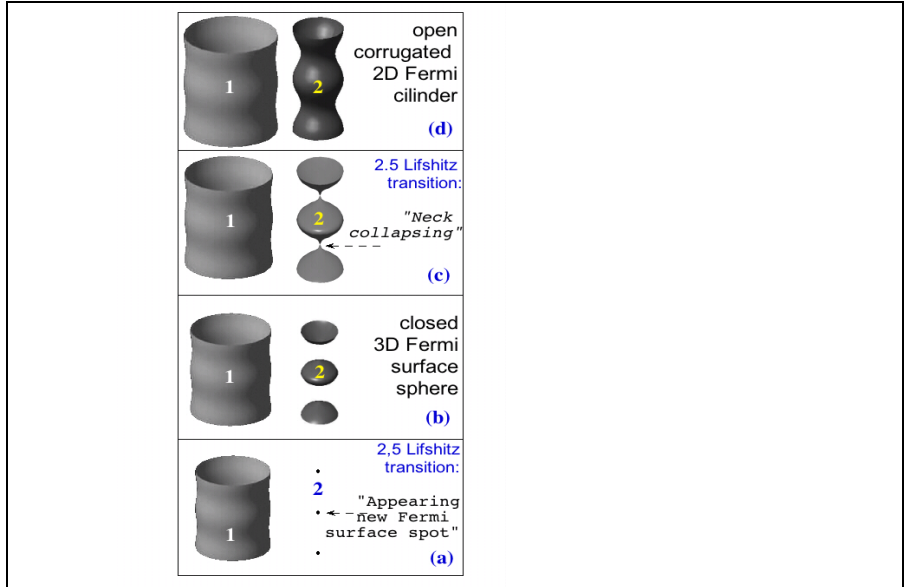

**Figure 3**: The evolution of Fermi surface topology in a generic two-band layered material where the chemical potential is moved near a band edge. A first type of 2.5 Lifshitz transition occurs here a new 3D Fermi surface appears (panel a). A second type of 2.5 Lifshitz transition occurs by moving further the chemical potential at the opening of a neck in a corrugated Fermi surface cylinder (panel c).

The presence of a 2.5 Lifshitz transition predicted also for cuprates in the two component scenario [27-30] was not accepted by the scientific community for many years but after 2003 many different experiments have observed 2.5 Lifshitz transitions in cuprates [31-37]. In cuprates a pseudo gap matter (probably made of polarons) where condensates should be in the antiadiabatic limit, was found to coexist with a Fermi liquid. The cuprates show beyond the insulator-to-metal transition two 2.5 Lifshitz transitions at doping 1/8 and near optimum doping as shown by the evolution of Hall effect [31] and of transport properties of the normal phase in high magnetic field [32,33], and quantum oscillations [35].

The presence of multiband superconductivity in proximity of band edges clearly appear in electron doped iron based superconductors in Ba(Fe$_{1-x}$Co$_x$)$_2$As$_2$ [21] and a(Fe$_{1-x}$Ni$_x$)$_2$As$_2$ [22] where the T$_c$ dome occurs in the proximity of 2.5 Lifshitz transitions [22] as predicted by Innocenti et al [20].

## 2. The multiband anisotropic BCS theory

Multiband superconductivity [38,39] appears in anisotropic superconductivity, where the gaps are different in different large Fermi surfaces. The multiple Fermi surfaces cannot be reduced to a single effective band since the bands have different symmetry and/or are located in different spatial portions of the material. In this scenario the single particle hopping in presence of impurities and hybridization are forbidden. For example the condensate many body BCS wave-function for a two band superconductor made of a first *a*-band and a second *b*-band is given by:

$$\left|\Psi_{Kondo}\right\rangle = \prod_{k}(u_k + v_k a^+_{k\uparrow}a^+_{-k\downarrow})\prod_{k'}(x_{k'} + y_{k'}b^+_{k'\uparrow}b^+_{-k'\downarrow})\left|0\right\rangle \quad (7)$$

The term corresponding to the transfer of a pair from the "a"-band to the "b"-band and vice-versa appears with the negative sign [40] in the expression of the energy.

$$\sum_{k,k'} J(k,k')(a^+_{k\uparrow}a^+_{-k\downarrow}b_{-k\downarrow}b_{k\uparrow}) \quad (8)$$

where $a^+$ and $b^+$ are creation operators of electrons in the "a" and "b" band respectively and $J(k,k')$ is an exchange-like integral. This gain of energy is the origin of the increase of the transition temperature driven by this exchange-like [40] interaction between pairs in different points in the k-space. This is called also a Josephson-like pairing term, in fact the Josephson effect [41] describes the transfer of pairs between two different condensates in different spatial positions. This Josephson-like pairing interaction appears in the standard BCS multiband theory with a square exponent therefore it may be repulsive as it was first noticed by Kondo [40]. It is therefore different from the Cooper pairing process that is the conventional attractive intraband attraction for electron at the Fermi level in the single band BCS theory. In the case of repulsive Josephson-like interaction in multiband superconductivity the order parameter shows the sign reversal between the different bands or different points in the k-space. This well established theoretical result became popular in 2008 since it was proposed for the case of iron based superconductors with the name of $s\pm$ mechanism.

The k-dependent gap in each band n depends on the gaps in other bands in multi-condensate superconductivity

$$\Delta_n(\mu, k_y) = -\frac{1}{2N} \sum_{n',k'_y,k'_x} \frac{V_{n,n'}(k,k')\Delta_{n'}(k'_y)}{\sqrt{(E_{n'}(k'_y) + \varepsilon_{k'_x} - \mu)^2 + \Delta^2_{n'}(k'_y)}} \quad (9)$$

where the k-dependent coupling is given by

$$V^o_{n,k_y;n',k'_y} = -V_o \int_S dx\,dy\, \psi_{n,-k}(x,y)\psi_{n',-k'}(x,y)\psi_{n,k}(x,y)\psi_{n',k'}(x,y) \quad (10)$$

**3. Shape resonance in superconducting gaps: a type of Fano resonance.**

In the standard theory of multiband superconductivity the Fermi level is assumed to far away from each band edge: i.e., the Fermi surfaces are large and intraband pairing in each band is in the adiabatic limit [38,39]. On the contrary some authors have proposed the idea of having a very small Fermi surface coexisting a large Fermi surface. The idea of having strong localized pairs coexisting with delocalized Copper pairs is quite old and goes back to the Bose-Fermi model [42-43] and the spin gap proximity effect mechanism of S. Kivelson, et al. [44].

The shape resonance model on the contrary considers the evolution of the Fermi surfaces tuning of the Fermi level from below (above) the bottom (the top) of a band to well above the band edge of a second band with symmetry different from the large first one [46]. Therefore the shape resonance theory considers the evolution of superconductivity from the BEC like regime to the two bands BCS regime where both Fermi surfaces are large. The shape resonance theory [13-20,46] assumes that the anisotropic BCS theory with no approximations works in agreement with Leggett [11].

Spectacular effects similar to a Fano resonance [47-52] occur in the superconductivity mechanism in a multiband superconductor when the chemical potential is tuned near a band edge at a 2.5 Lifshitz transition. In this case the intraband pairing mechanism enters in the antiadiabatic limit in the particular band where the Fermi level is near the band edge.

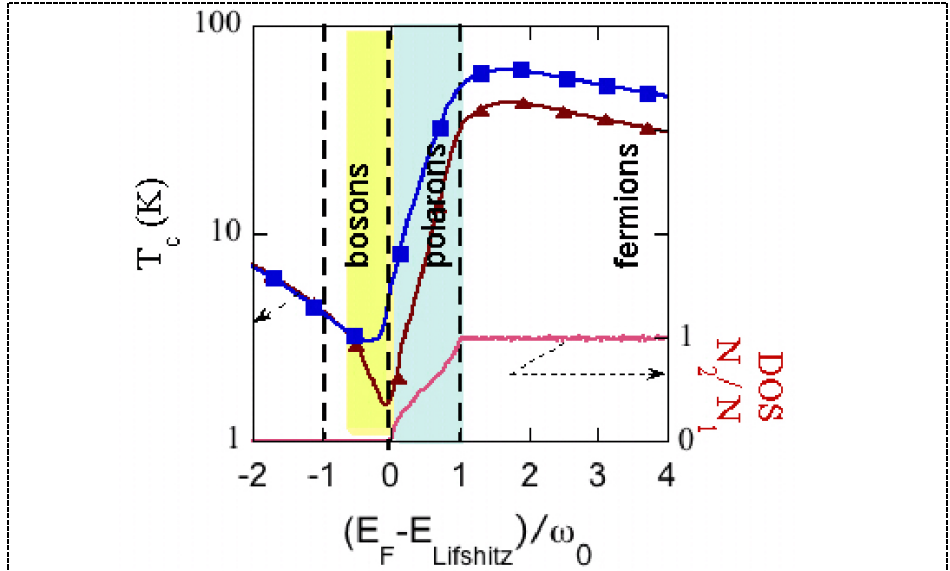

**Figure 4**: The critical temperature of a layered multi-band superconductor tuning the chemical potential near a 2.5 Lifshitz transition where the Fermi surface of a new appearing band evolves as shown in Fig. 3. The critical temperature in log-scale is plotted as a function of the Lifshitz energy parameter $z = (E_F - E_{Lifshitz})/\omega_0$ where $\omega_0$ is the energy of the attractive intraband pairing interaction. The plot show two cases of Josephson-like pair exchange coupling ratio: a) $c_{12}/c_{11}$=-1.08 (*blue filled squares*) and $c_{12}/c_{11}$=-0.43 (*black filled triangles*). The colored regions indicate the BEC regime of the condensate (yellow) -1<z<0 and bipolaronic condensate (blue) 0<z<1 in the new appearing Fermi surface.

The system can be treated in the BCS scheme [11] if the equation for the critical temperature near a 2.5 Lifshitz transition is solved together with the density equation as suggested by Leggett [5] taking into account that in this regime there is a large variation between the chemical potential in the superconducting phase and the normal phase

$$\Delta_n(k) = -\frac{1}{N}\sum_{n'k'} V_{n,n'}(k,k') \frac{tgh(\frac{\xi_{n'}(k')}{2T_c})}{2\xi_{n'}(k')}\Delta_{n'}(k') \qquad (11)$$

$$\rho = \frac{1}{S}\sum_{n}^{N_b}\sum_{k_x,k_y}(1 - \frac{\varepsilon_n(k_x,k_y) - \mu}{\sqrt{(\varepsilon_n(k_x,k_y) - \mu)^2 + \Delta_{n,k_y}^2}}) \qquad (12)$$

We consider a two band superconductor where there is first large 2D cylindrical Fermi surface and the chemical potential is tuned across a 2.5 Lifshitz transition so that the Fermi surface topology of the new appearing Fermi surface evolves as shown in Fig. 2. The superconducting critical temperature as a function of the Lifshitz energy parameter $z = (E_F - E_{Lifshitz})/\omega_0$ where $\omega_0$ is the energy of the attractive intraband pairing interaction. measuring the energy distance form the energy $E_{Lifshitz}$ of the 2.5 electronic topological Lifshitz transition for the appearing of the a new Fermi surface as plotted in Fig. 4. This figure provides a typical example of shape resonance in the superconducting gaps.

The chemical potential crosses the first 2.5 Lifshitz transition, for the appearing of the new Fermi surface spot, at the value of the Lifshitz energy parameter $z = 0$, and the second 2.5 Lifshitz transition "opening a neck" at $z = 1$. The coupling in the first Fermi band is assumed to be in the standard BCS weak coupling regime $c_{11} = 0.22$. We consider here the case of strong coupling in the second band where the attractive coupling term $c_{22}$ is about two times larger than the Cooper pairing coupling parameter in the first band $c_{22}/c_{11} = 2.17$. When the Lifshitz energy parameter is in the range $-1<z<0$ a *BEC condensate* is formed in the second band since all charges in the second band condense. At $z = 0$, we show evidence for the antiresonance due to negative interference effects between different pairing channels in the BCS condensate in the first band and the BEC in the new appearing band. This clearly shows is the quantum interference nature of shape resonances. The second curve shows the higher critical temperature for a larger $c_{12}$ Josephson-like pair transfer value. $T_c$ reaches the minimum value at the antiresonance for $z = 0$ in the limit of weak Josephson-like coupling, while for a strong Josephson-like coupling $c_{12}$, $c_{21}$ the minimum due to the Fano-like antiresonance approach $z=-1$. Clearly we show in Fig. 4 that the shape resonance has a similar line-shape as a Fano resonance. Our minima correspond to the zero crossing which occurs near a Feshbach resonance in the ultracold Fermi gas problem.

These results provide a roadmap for material design of new room temperature superconductors with a lattice geometry that allow :

a) a multi-band superconductor in the clean limit with several bands crossing the Fermi level;
b) the different bands crossing the Fermi energy should have *different parity* and *different spatial locations* to avoid hybridization
c) the chemical potential is tuned near a 2.5 Lifshitz transition near a band edge.
d) the exchange-like or Josephson-like pair transfer integral should be as large as possible
e) a first condensate is in the adiabatic regime and a second condensate in the anti-adiabatic regime
f) The chemical potential can be tuned by pressure or gate voltage to get the Fano shape resonance in the superconducting gaps between pairing scattering channels in a BCS-BEC crossover
g) Tune the chemical potential by gate voltage tuning of using illumination or pressure so that the bipolaron condensate is converted into a BCS condensate in fact the maximum $T_c$ occurs at a Lifshitz energy parameter z near 1.5 -2.

## 4. Optimum lattice inhomogeneity favoring high $T_c$.

The metals near a 2.5 Lifshitz transition are in a metastable phase in the verge of phase separation in the presence of interactions. It has been found that the metastability of the lattice form a network of nanoscale striped puddles of lattice, charge and orbital density waves called "superstripes" [53-56]. In each puddle the lattice and/or charge and/or spin and/or orbital 1D modulation gives multiple subbands crossing the Fermi level resulting in multiple Fermi surfaces in heterogeneous spots in the k-space.

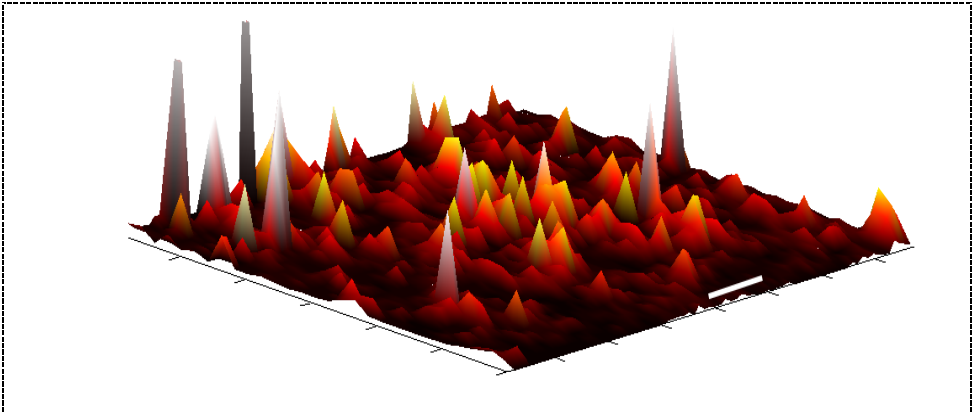

**Figure 5**: The landscape of nanoscale striped puddles of "local lattice distortions" in a cuprate superconductor $La_2CuO_{4+y}$ measured by scanning nano x-ray diffraction [57] realizing the "superstripes" scenario [53-56]. The position dependence of the Q3-LLD superstructure intensity in the $La_2CuO_{4+y}$ crystals with critical temperature, $T_c$, 37 K is plotted. The white bar indicates a length of 100 microns.

Indeed, in samples where a the dopants are frozen in a random distribution at a temperature lower than 200K, the local lattice distortions (LLD) detected by EXAFS XANES [29,30] and diffuse x-ray scattering show the formation of local lattice distortions that get self organized forming networks of nanoscale (10-100 nm) striped puddles [53-56] that recently have been visualized by scanning nano x-ray diffraction [57] shown in Fig. 5. The scanning XRD images show that a better self organization of LLD droplets favors higher $T_c$. In particular the density distribution of the nano-puddles follow a power-law distribution with a cut off. Investigating several sample it emerges that increasing the cut-off, i.e., going from an exponential to power law distribution the critical temperature increases as predicted by a recent theory for granular superconductors made of networks with power law distribution of superconducting nano grains connected with Josephson-like links.

Using a x-ray nanosized beam of 300 nm in diameter, we have obtained the imaging of the regions in $La_2CuO_{4+y}$, that contain incommensurate modulated local lattice distortions (LLD) shown in Fig. 5. The puddles are determined by self organization of local lattice distortions with superlattice wave-vector **q3** = 0.21 **b***+0.29 **c** therefore are called with the acronym Q3-LLD [57]. Fig. 4 shows the spatial distribution of Q3-LLD puddles for a superconducting sample. The Q3-LLD droplets form networks whose nature varies with superconducting critical temperature. We have used X-ray micro-diffraction apparatus at the ESRF to map the evolution of the Q3-LLD satellites for four single crystals of electrochemically doped $La_2CuO_{4+y}$, from the underdoped state to the optimum doping range, $0.06<y<0.12$.

The probability distributions, of the Q3-LLD XRD intensity for single crystals of electrochemically doped $La_2CuO_{4+y}$ from the underdoped state to the optimum doping range, $0.06 < y < 0.12$ follow a power law distribution $P(x) \propto x^{-\alpha} \exp(-x/x_0)$ with a variable exponential cut-off $x_0$ with a constant power-law exponent $\alpha = 2.6 \pm 0.1$.

In Fig. 6 we have plotted the critical temperature of samples in the range 27-38 K associated with the droplet network of Josephson coupled nano-grains, as a function of the cut-off of the probability distribution $x_0$ of the intensity of x-ray intensities due to the variable density of Q3-LLD puddles. The critical temperature scales with the cut-off according to a power law with an exponent $0.4 \pm 0.05$. This result points again toward the importance of connectivity and an optimum inhomogeneity for high critical temperature. They are in qualitative agreement with the theoretical prediction of the increase of $T_c$ in a granular superconductor on an annealed complex network made of Josephson coupled grain following a power law distribution with a finite cut-off [57]. In fact, for a power law distribution of links in a granular superconductor with an exponent $\alpha=2.6$, the critical temperature is predicted to increase as a function of the cut-off with an exponent $3-\alpha$, as observed experimentally supporting the theory of quantum phase transitions in network of superconducting grains [58,59] and for bosonic scale free networks [60].

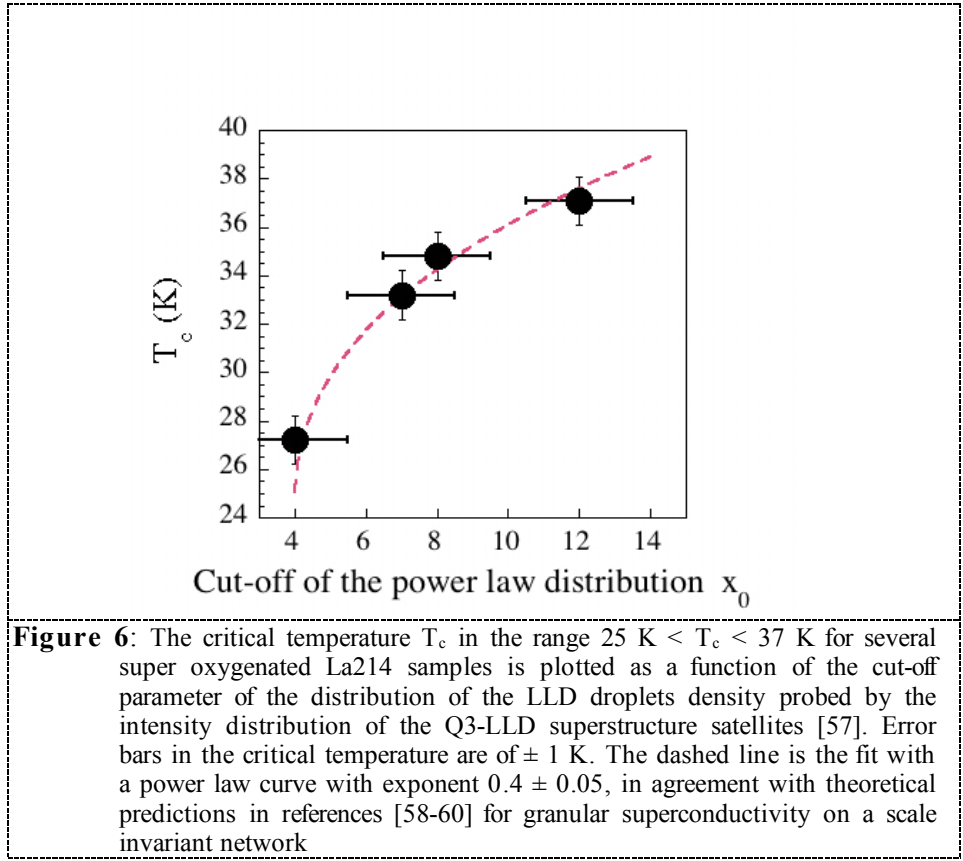

**Figure 6**: The critical temperature $T_c$ in the range 25 K $< T_c <$ 37 K for several super oxygenated La214 samples is plotted as a function of the cut-off parameter of the distribution of the LLD droplets density probed by the intensity distribution of the Q3-LLD superstructure satellites [57]. Error bars in the critical temperature are of $\pm 1$ K. The dashed line is the fit with a power law curve with exponent $0.4 \pm 0.05$, in agreement with theoretical predictions in references [58-60] for granular superconductivity on a scale invariant network

In the superlattices the lattice complexity appears not only in the superconducting planes but also in the spacer layers. There the defect self organization forming striped structure will induce Fermi surface reconstruction in the superconducting layers that controls the Fermi surface topology so it can shift the 2.5 Lifshitz transitions.

The oxygen interstitials can be very inhomogeneous, even in "optimal" superconducting samples with a scale free distribution favoring high $T_c$ as recently shown by scanning micro

X-ray diffraction [61] and the oxygen interstitials self organization can be controlled by x-ray illumination [62-64].

The observed oxygen interstitals mobility is controlled by the tensile microstrain in the spacer layers due to the lattice misfit strain between the different units forming the super-lattices in Fig. 2 [65-69]. It is possible that the maximum Tc of 160K at a particular compressive 2% microstrain in cuprates is related with optimum self organization of the nanoscale puddles approaching the scale free distribution. Many authors are now reaching the conclusion that an optimum inhomogeneity in complex matter favors high temperature superconductivity [70]. This idea is supported by the evidence that a similar nanoscale phase separation has been observed in iron based superconductors [71-72]. The emerging picture is that the first-order transition appears in the proximity of a 2.5 Lifshitz transit when the electron Coulomb interaction is switched on and recently interaction effects on the Lifshitz transitions have been systematically studied [73-75] while the topological change in the Fermi surface (called the 2.5 Lifshitz transition) was originally studied for non interacting electrons. The frustrated or arrested phase separation that appears in the experiments probing the structure fluctuations beyond the average structure point toward a key role of polaronic charge carriers in small Fermi surface [76].

## 5. Conclusions

We have presented the superstripes scenario for high temperature superconductors. These complex materials show k-dependent anisotropic pairing, superconductivity in the clean limit and multi-condensate superconductivity. The high $T_c$ dome appears near a 2.5 Lifshitz transition near a band edge. The electrons in one of the multiple Fermi surfaces are in the BEC-BCS crossover regime. At list one of the borders of the high $T_c$ dome is determined by the Fano anti-resonance of the shape resonance in the superconducting gaps where $T_c$ drops toward zero at the 2.5 Lifshitz transition for the appearing of a new Fermi surface. At this particular regime there is a BEC-BCS crossover with coexisting BEC and BCS condensates that gives an anti-resonance in the configuration interaction controlled by the Josephson-like pair transfer term. This term gives an increasing high Tc amplification when the BEC is converted to a polaronic pair condensate and finally the maximum $T_c$ is reached where the polaronic condensate moves toward a BCS condensate.

The materials near a 2.5 Lifshitz transition are in a metastable phase due to lattice instability and show networks of striped superconducting grains. It seem to us that the maximum $T_c$ is reached at the particular condition where in each grain with a striped structure the critical temperature is amplified by the shape resonance in the superconducting gaps and the grains form a scale free network where $T_c$ is pushed to the highest possible temperature.

**Acknowledgments:**
The author thanks Tony Leggett, Lev Gorkov, Andrea Perali, Carlos Sa de Melo, Klim Kugel, Masatoshi Imada for useful discussions.